\documentclass[a4paper]{article}
\usepackage{a4wide}
\usepackage{amsmath,amsfonts,amssymb,amsthm,graphicx}
\newcommand{\Nat}{\mathbb {N}}            
\newcommand{\Int}{\mathbb {Z}}            

\DeclareMathOperator*{\slim}{{\it s}-lim}

\DeclareMathOperator{\imag}{Im}

\newtheorem{theorem}{Theorem}[section]

\newtheorem{corollary}[theorem]{Corollary}
\newtheorem{lemma}[theorem]{Lemma}

\newtheorem{conjecture}[theorem]{Conjecture}
\theoremstyle{definition}
\newtheorem{remark}[theorem]{Remark}

\numberwithin{equation}{section}

\usepackage{color}

\begin{document}


\vspace*{1cm}

\begin{center}\large\bf
A partition-free approach to transient and steady-state charge currents
\end{center}

\begin{center}
\small{
Horia D. Cornean\footnote{Department of Mathematical Sciences,
    Aalborg
    University, Fredrik Bajers Vej 7G, 9220 Aalborg, Denmark; e-mail:
    cornean@math.aau.dk},
C{\'e}line Gianesello \footnote{Universit{\'e} du sud Toulon-Var \& Centre de
Physique Th\'eorique, Campus de Luminy, Case 907
13288 Marseille cedex 9, France; e-mail: gianesello@cpt.univ-mrs.fr  },
Valentin Zagrebnov \footnote{Universit{\'e} de la M{\'e}diterran{\'e}e
  \& Centre de Physique Th\'eorique, Campus de Luminy, Case 907
13288 Marseille cedex 9, France; e-mail: Valentin.Zagrebnov@cpt.univ-mrs.fr}}

\end{center}

\vspace{0.5cm}

\noindent

\begin{abstract}
\noindent {We construct a non-equilibrium steady state and calculate
the corresponding current for a
mesoscopic Fermi system in the \textit{partition-free} setting. To this end we
study a small sample coupled to a finite
number of semi-infinite leads.} Initially, the whole system of
quasi-free fermions is in a grand canonical equilibrium
state. At $t=0$ we turn on a potential bias on the leads and let the
system evolve. We study how the charge
current behaves in time and how it stabilizes itself around a steady
state value, which is given by a
Landauer-type formula.
\end{abstract}

\vspace{0.1cm}

\section{Introduction}

{At the present time one can essentially distinguish two different
ways of constructing non equilibrium steady states (NESS) for composed systems.

The first method consists of preparing a \textit{partitioned} initial
state for the total system containing several sub-systems, each of
which being in a different state of thermal
equilibrium, and then put
them into contact with each other at $t=0$, and let the coupled total
system evolve in time until it reaches a steady state. In the
mathematical physics community this method goes back to D. Ruelle \cite{Ru1},
\cite{Ru2}. It was seriously promoted during the recent years
through numerous papers, see e.g.
\cite{AtJoPi, AJPP, JP1, Ru2, N, brarob}
and references therein.  One can allow the carriers to interact in the
sample \cite{JOP}, and the theory
still works. Note that even if one chooses to turn on the coupling
between the reservoirs in a time dependent way (for example
\textit{adiabatically}), the results remain the same \cite{CNZ}.}

The second method deals with those situations in which the initial state is
an equilibrium state for the already coupled  (i.e. \textit{partition-free})
total system. {The partition-free approach goes back at least to M.Cini \cite{Cini}.
This means that the initial state is not "partitioned" into a direct sum of
equilibrium sub-states associated to e.g. different leads.
The system is taken out of equilibrium by switching on
an electrical bias between subsystems (leads) like for example
turning on a d.c. battery,
which in a certain way can be seen as changing the electro-chemical
potentials of the leads coupled via a small
sample. In contrast to the first method, there are almost no rigorous
mathematical results on the second method beyond the linear response
theory, or at least we are not
aware of the existence of such results.}

Although these two methods seem very similar, especially if one
suddenly switches on the parameter bias in the
partition-free system at $t=0$, their implementations are different.
One of the aims of this paper is to illustrate this observation.

The main result of the present paper is that now we are able to
construct a NESS and to study charge
currents in the partition-free setting and for the full response.
Let us describe in words what we do.

For simplicity, in this paper we only consider
two semi-infinite leads which are both
coupled with the same small sample when $t < 0$. The full system is in a
Gibbs equilibrium state at a given temperature and
chemical potential. At $t=0$ we turn on a time-dependent potential
bias $V(t)$ between the leads, modeling a
transient regime of a d.c. battery. At time $t_1>0$ the bias is
stabilized and remains constant in time afterwards.
The statistical density matrix $\rho(t)$ is found as the solution of
a quantum Liouville equation, with an initial condition given by our
global Gibbs state at $t=0$. The time-dependent charge current from
one lead to the other is defined as the mean
value of a current operator in the state $\rho(t)$, see
\eqref{atreia1}-\eqref{acincea1} for
details. A priori the current depends on time, on the way
we switch on the
bias, and the point where we make the measurement.

In Theorem \ref{teorema1} we show the existence and
compute the ergodic limit of this charge current. The limit depends
neither on the way we switch on the bias, nor on the point where we
measure the current. We also obtain an explicit Landauer-type formula for
this limiting charge current value, involving the transmission
coefficients between the leads.

Establishing Landauer-B\"uttiker type formulas (see e.g.
\cite{But, BPT, Avron1, Avron2}) starting from first
principles but in the partition free setting was the original
motivation of a number of remarkable physical papers,
see for example \cite{FL}, \cite{LA}, \cite{BS} and references
therein. Probably the state of the art of this subject seen from a
physical perspective is to be found in two papers by Stefanucci
and collaborators \cite{Stefa1, Stefa2} in which
the partition free approach is combined with the Green-Keldysh theory
and a number of very current interesting formulas are proposed.

A first mathematically sound derivation of the Landauer-B\"uttiker
formula in the partition free approach under the linear response
approximation was obtained in
\cite{CJM1} and further investigated in \cite{CJM2}. In \cite{CDNP-1}
we significantly improved the method of proof
of \cite{CJM1}, which also allowed us to extend the results to the
continuous case. Another challenging open
problem is to extend the formalism in order to accommodate more
efficient numerical current computations in
transient regimes (see \cite{MGM1, MGM2, MGM3, Stefa3} and references
therein), and locally interacting fermions.

The structure of the rest of the paper is the following:
\begin{itemize}

\item In Section \ref{themodel} we introduce the model and define the transient
charge current in \eqref{acincea1}. The main result is formulated in
Theorem \ref{teorema1}.

\item Section \ref{maintheorem} starts with a list of well-known facts
  about the spectral and scattering theory of mesoscopic systems
  coupled to semi-infinite leads. The second part of the section is
  dedicated to the proof of our main
theorem. At the end we give a list of open problems.
\end{itemize}

\section{Set up and main results}\label{themodel}

\begin{figure}[h]
\begin{center}
\includegraphics[scale=0.3]{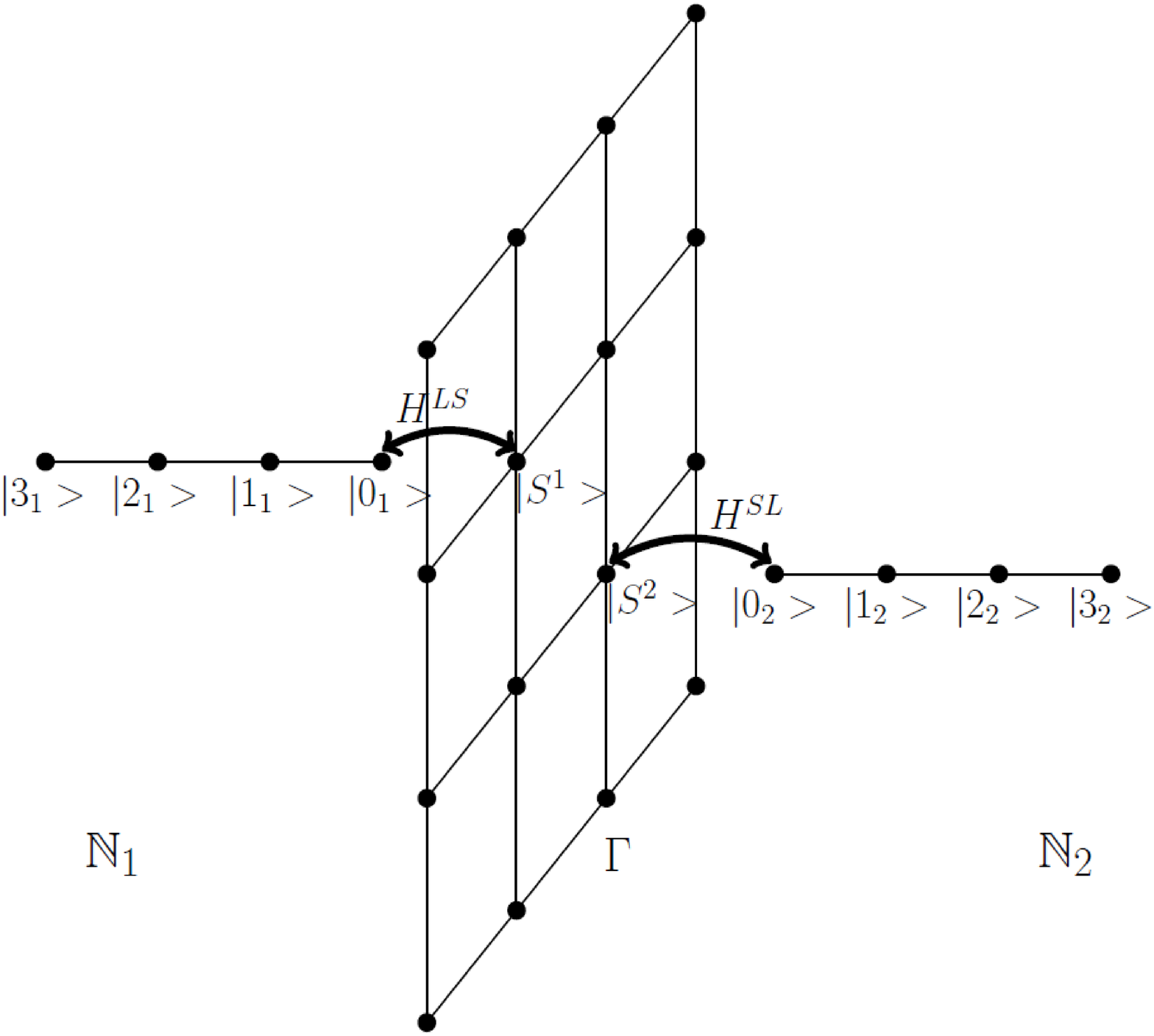}
   \vspace{1cm}
   \caption{A sample $\Gamma$ connected to two semi-infinite leads $\mathbb{N}_1$ and $\mathbb{N}_2$.
   Here $|j_\alpha\rangle$  denotes a basis element at the site with the number $j$ of the 
   lead $\alpha$}
\end{center}
\end{figure}

We work with a discrete model in a one-particle Hilbert space $\cal
H$. Following the physical convention we
define the scalar product to be linear with respect to the second variable.

Our carriers are quasi-free fermions (electrons). A small sample $S$
is modeled by $\Gamma \subset \mathbb{Z}^{2}$, chosen to be a finite
subset of $\Int^2$. We couple $S$ to two "one-dimensional" semi-infinite discrete leads
$\alpha =1,2$. The sites of a lead (building its standard
basis) are indexed by the set $\mathbb{N_\alpha}:=\{0,1,2,...\}$.
Thus, $|j_\alpha\rangle$  denotes the basis element at
the site with the number $j$ of the lead $\alpha$, see Fig.1.
The total one-particle Hilbert space is a direct sum of the space
modeling the sample $\Gamma\subset \mathbb{Z}^{2}$, and two spaces corresponding to
the leads $\{\mathbb{N_\alpha}\}_{\alpha =1,2}$ :
\begin{equation}\label{tothilbs}
{\cal H}:=l^2(\Gamma)\oplus l^2(\mathbb{N}_{1})\oplus l^2(\mathbb{N}_{2}) \ .
\end{equation}
We denote by $\{|m,n\rangle\}_{(m,n)\in\mathbb{Z}^2}$ and by
$\{|j_{\alpha}\rangle\}_{j_{\alpha}\in\mathbb{N}_\alpha}$ the corresponding orthonormal bases
of the spaces $l^2(\mathbb{Z}^{2})$ and $l^2(\mathbb{N}_{\alpha})$, where $\alpha =1,2$.

Now we describe our one-particle Hamiltonian. For the sample $S$ we may
choose any self-adjoint bounded operator $H^S$. For example, we can choose $H^S$ to be the
restriction to $l^2(\Gamma)$, of a lattice Harper-type operator with
Dirichlet boundary conditions on $\Gamma$, but the concrete model for
$H^S$ does not play any role in the proof of our results. Notice that $\Gamma$ is chosen
to be finite, but can be arbitrarily large.

On each lead $\alpha =1,2$ we define the identical one-dimensional discrete Laplacians
acting on the functions from $l^2(\mathbb{N}_{\alpha})$ with Dirichlet
boundary conditions on $\mathbb{N}_\alpha$:
\begin{align}\label{hasl}
& (H^L_{\alpha}\Psi)(n):=
t_L  \left\{\Psi(n+1)  +\Psi(n-1)\right\},\quad n\geq 1;\quad
(H^L_{\alpha}\Psi)(0):=t_L  \Psi(1),\nonumber \\
&H^L:=\sum_{\alpha=1}^2 H^L_{\alpha},
\end{align}
where $t_L>0$ is a hopping constant. In the following, we denote the
Hamiltonian corresponding to these three \textit{disconnected} subsystems by:
\begin{equation*}
H_0:=H^L+H^S.
\end{equation*}
The coupling between the sample and leads is described by the \textit{tunneling} Hamiltonian
(see Fig.1):
\begin{equation}\label{hast}
H^T:=\tau\sum_{\alpha=1}^2\{|0_{\alpha}\rangle\langle {\cal S}^\alpha |
+|{\cal S}^\alpha\rangle\langle 0_{\alpha}|\}
=: H^{LS}+H^{SL}.
\end{equation}
Here $\tau>0$ is the hopping parameter between leads and the
sample. The interaction (\ref{hast}) simulates
a quantum point constriction, or a tunneling barrier. Here
$|0_{\alpha}\rangle$ is the \textit{first} site on
the lead $\alpha$, and $|{\cal S}^\alpha\rangle$ is
the corresponding \textit{contact} site $|m_{\alpha},n_{\alpha}\rangle\ $ on the sample coupled to the
lead $\alpha$.

Then the total one-particle Hamiltonian takes the form:
\begin{equation}\label{hastot}
H:=H^S+\sum_{\alpha=1}^2 H^L_{\alpha}+H^T=H^S+H^L+H^{LS}+H^{SL}{=:H_0+H^T}.
\end{equation}

\begin{remark}
As we mentioned before, our results can be extended through verbatim to more general
choices of Hamiltonians $H^L$ and $H^T$. The key properties that we need are: the \textit{absolutely
continuous} spectrum in the leads, and a \textit{finite rank} operator
coupling between a finite sample and
the leads.
\end{remark}

\subsection{The state and charge current}

At $t<0$ the total coupled system (\ref{hastot}) is at equilibrium for a given temperature
$1/\beta \geq 0$ and a
chemical potential $\mu$. Since we work with non-interacting fermions,
the corresponding one-particle
\textit{Fermi-Dirac} equilibrium density matrix is the operator
\begin{equation}\label{FD}
f(H)=\frac{1}{e^{\beta(H-\mu)}+1} \ , \ \mu\in\mathbb{R}\ ,
\end{equation}
defined on the Hilbert space (\ref{tothilbs}).

At the moment $t=0$ we turn on a bias on lead number \textit{one} in
the following way. We fix $t_1>0$ and
choose a real and continuous function $\phi$ which has the property that
$\phi(t)=0$ if $t<0$ and $\phi(t)=1$ if $t>t_1$. Let $v>0$. Denote by
$P_1: \mathcal{H} \mapsto l^2({\Nat}_1)$ the projection on the lead number \textit{one}.
Then define the time dependent potential bias as:
\begin{equation}\label{prima1}
V_1(t):=v\phi(t)P_1 \ .
\end{equation}
Denote by $U(t)$ the unitary evolution associated to $H+V_1(t)$ through
the time dependent Schr\"{o}dinger equation:
\begin{align}\label{adoua1}
i \partial_{t}U(t)=(H+V_1(t))U(t),\quad U(0)= \mathbb{I} \ .
\end{align}
The density matrix at time $t>0$ is a solution of the Liouville equation and
be expressed by:
\begin{align}\label{atreia1}
\rho(t):=U(t)f(H)U(t)^* \ .
\end{align}
Denote by $P_2^{(n)}: \mathcal{H} \mapsto l^2(\{n,n+1,\ldots\})$,
the projection on the second lead from which we exclude the first $n$
sites. If $n=0$, then it is just the projection $P_2$ on the lead 2. We define the
current operator modeling the measurement of the charge flow at site $n$ by:
\begin{align}\label{apatra1}
j_n:=i[H+V_1(t),P_2^{(n)}]=i[H,P_2^{(n)}] \ , \quad j_0=i[H^T,P_2] \ .
\end{align}
\begin{remark}
Clearly, the current operator has finite rank, thus it is trace class. This
is one important feature which is only true in the \textit{discrete}
setting. It significantly simplifies the technical estimates compared
to the \textit{continuous} case.
\end{remark}
\begin{remark}\label{supp-P}
To obtain (\ref{apatra1}) we used some evident support properties of
the projections $P_2^{(n)}$, which imply:
\begin{equation}\label{comm1-P}
[P_1, P_2^{(n)}] =0, \quad P_2^{(n)}P_2^{(n+1)}=P_2^{(n+1)}, \quad
[H^{S}, P_2^{(n)}] =0,\quad \forall n\geq 0,
\end{equation}
and also:
\begin{align}\label{comm2-P}
[H^{L}, P_{2}] =0 \ , \ [H^{L}, P_{2}^{(n)}] \neq 0, \quad [H^{T}, P_2^{(n)}] =0 \quad \forall n\geq 1.
\end{align}
These properties make the commutators in (\ref{apatra1}) nontrivial,
and are important for the study of the
current propagation (see the proof of the point {\it (ii)} of our main theorem).
\end{remark}
The charge current flowing through the second lead at time $t>0$ and
measured at site $n$ is the expectation of
the operator $j_n$ from (\ref{apatra1}) in the quasi-free state
defined by the time dependent one-particle density matrix $\rho(t)$:
\begin{align}\label{acincea1}
I(t,n):={\rm Tr} \{\rho(t)j_n\}.
\end{align}

\subsection{The main theorem}

Now we are ready to formulate our main results, collected in one theorem.
\begin{theorem}\label{teorema1}
\noindent {\rm (i)} The following ergodic limit exists and is
independent of $n$, $t_1$ and $\phi$:
\begin{equation}\label{asasea1}
I_\infty:=\lim_{T\to \infty}\frac{1}{T}\int_0^T I(t,n)dt.
\end{equation}

\noindent {\rm (ii)} Fix $t\geq 0$. Then the current vanishes if we measure it infinitely far inside
the lead 2:
\begin{equation}\label{martie101}
\lim_{n\to \infty} I(t,n)=0.
\end{equation}

\noindent {\rm (iii)} Assume that the operator $H+vP_1$ has only
finitely many eigenvalues. If we measure the current very far inside the
second lead (but not infinitely far), the transient$/$oscillatory effects
will be weaker and weaker and the current defined in
(\ref{acincea1})
will slightly fluctuate around the value (\ref{asasea1}). More precisely:
\begin{equation}\label{asaptea1}
\lim_{n\to\infty}\limsup_{t\to \infty}|I(t,n)-I_\infty|=0.
\end{equation}

\noindent {\rm (iv)} Denote by ${\cal T}_{12}^{(v)}(\lambda)$ the
transmittance coefficient between the two leads at bias $v$ for the spectral parameter $\lambda$
{\rm{(}}see \eqref{transmys} for a rigorous definition{\rm{)}}. Then we establish the following
Landauer-type formula:
\begin{equation}\label{prima5}
 I_\infty=2\pi\int_{[-2t_L+v,2t_L+v]\cap [-2t_L,2t_L]}\left \{f(\lambda)-f(\lambda-v)\right \}
{\cal T}_{12}^{(v)}(\lambda)d\lambda \ .
\end{equation}
\end{theorem}

\begin{remark}\label{asymptotic current}
From {\it (ii)} one concludes that if we fix the time of measurement
$t$ and push the measuring point $n$ to infinity, the
current tends to zero. Although this does not imply that the propagation speed
of the current is \textit{finite}, a property which we cannot expect
to hold true because of the non-relativistic dynamics on the leads.
On the other hand, in {\it (iii)} we prove that if the current
measuring device is placed further and further away from the sample,
after waiting a very long time the
current becomes non-zero and has weaker and weaker fluctuations around its steady-state
mean value. In other words, the limits ${t\to\infty}$ and
${n\to\infty}$ do not commute.
\end{remark}

\begin{conjecture}\label{Lieb-Robinson}
{As a complement to Remark \ref{asymptotic current}, we conjecture
that the \textit{group-velocity} of the spatial
correlations in our model is finite, i.e. if $A$ is any observable
supported in a neighborhood of the sample, then there exist some
positive constants $C$, $M$ and $\nu$ such that the Lieb-Robinson type
bound {\rm{\cite{LR}:}}
\begin{equation}\label{LR}
\|[A, U(t)^* P_2^{(n)}U(t)]\| \leq C e^{-M(n -\nu t)}
\end{equation}
holds true for every $t>0$ and $n\geq 0$. The exponential bound sounds as
a strong one, i.e. one can not exclude {\rm{a priori}} a power-like decay.}
\end{conjecture}

\begin{remark}
Our proofs are exclusively based on one-body scattering methods.
We do not use the many-body language, which is unavoidable only if
the carriers interact.
\end{remark}

\begin{remark}
By the same reasons, one can completely characterize the many-body
states $\omega (\cdot)$ on the Fermi
algebra CAR($\mathcal{H}$), (algebra of the Canonical Anticommutation
Relations) by a one-particle density-matrix
operator $\rho$ defined on $\mathcal{H}$. If at $t=0$ the state
$\omega (\cdot)$ is the grand-canonical
equilibrium state on CAR($\mathcal{H}$) of a non-interacting Fermi system (\ref{hastot})
(equilibrium quasi-free state), then the density-matrix operator is
equal to $\rho(t=0) = f(H)$ (\ref{FD}).
The evolution (\ref{adoua1}) preserves this property, i.e. it
transforms this state into a non-equilibrium quasi-free state
(\ref{atreia1}).
\end{remark}

\section{Proof of the main theorem}\label{maintheorem}

We start this section with a list of well known facts about the spectral and scattering theory
of mesoscopic, systems coupled to semi-infinite leads. This will help us to fix notation and
streamline the proof of the theorem.

\subsection{Some spectral and scattering background}\label{scatteringtheory}

First we recall some elements of the stationary scattering problem
associated with the pair of Hamiltonians
$(H+vP_1,H_0+vP_1)$, where $H:=H_0 + H^T$. In this case the
free system consists of the leads
with a bias $v$ localized on the first lead together with the
\textit{decoupled} inner sample, and it is described by the
Hamiltonian $H_0+vP_1$. The perturbed system also contains the coupling $H^T$.

The operator $H_0=H^L+H^S$ has as a subspace of absolute continuity
$\mathcal{H}^{ac}(H_0)=\bigoplus_{\alpha=1}^2
l^2(\mathbb{N}_{\alpha})$. Since the operator $H-H_0$ is of finite
rank, the trace class scattering
theory implies that the M{\o}ller wave operators
\begin{equation}\label{omegapm}
\Omega_{\pm}^{(v)}=\slim_{t\to\mp\infty}e^{it(H+vP_1)}e^{-it(H_0+vP_1)}E_{\rm ac}(H_0+vP_1) \ ,
\end{equation}
exist and are complete, see e.g. \cite{RS3}, \cite{Yafaev}. Here $\quad E_{\rm ac}
(H_0+vP_1)= E_{\rm ac}(H_0)$ denotes the projection on the absolutely continuous subspace
$\mathcal{H}^{ac}(H_0+vP_1)\subset \mathcal{H}$, or $\mathcal{H}^{ac}(H_0)\subset \mathcal{H}$
of the corresponding operators. The location and nature of the spectrum of operators like
$H$ was extensively studied in \cite{CJM1}; one can prove under
generic conditions that there are only finitely many
eigenvalues, while the singular continuous spectrum is always absent.

It is known that the set of (normalized) generalized eigenfunctions of $H^L$ on the semi-infinite leads
$\alpha=1,2$ have the form:
\begin{equation}\label{apatra2}
\Psi_\alpha(\lambda)=\sum_{m\geq 0}\Psi(\lambda;m)
|m_\alpha\rangle,\quad \Psi(\lambda;m)=
\frac{\sin(k(m+1))}{\sqrt{\pi t_L\sin(k)}}.
\end{equation}
Here the spectral parameter $\lambda= \lambda_k(:=2t_L \cos(k))$ for
$k\in(-\pi,\pi)$. The generalized
Fourier transformation associated to these eigenvectors is defined by
\begin{gather}\label{genfour}
F\colon \bigoplus_{\alpha=1}^2 l^2(\mathbb{N_\alpha}) \to \bigoplus_{\alpha=1}^2
L^2([-2t_L,2t_L]), \\
[F(\Phi)]_\alpha(\lambda)=\langle
\Psi_\alpha(\lambda),\Phi_\alpha\rangle_{l^2(\mathbb{N}_{\alpha})}=
\sum_{m\geq 0}\overline{\Psi(\lambda;m)}\Phi_\alpha(m).
\label{genfour-b}
\end{gather}
Its adjoint is given by
\begin{gather}\label{genfouradj}
F^*\colon \bigoplus_{\alpha=1}^2
L^2([-2t_L,2t_L])\to \bigoplus_{\alpha=1}^2 l^2(\mathbb{N}_{\alpha}), \\
[F^*(\Xi)]_\alpha(m)
=\int_{-2t_L}^{2t_L}\Xi_\alpha(\lambda)\Psi(\lambda;m)d\lambda.
\label{genfouradj-b}
\end{gather}
We see that $F$ is a unitary operator, and that $FH^LF^*$ is just
the multiplication by $\lambda {\mathbb{I}}$ on the space which is a direct integral
$\int_{[-2t_L,2t_L]}^{\oplus}\mathbb{C}^2d\lambda \cong \bigoplus_{\alpha=1}^2
L^2([-2t_L,2t_L]) $, i.e.
\begin{equation}\label{jjjkkkk}
FH^LF^*\cong  \int_{[-2t_L,2t_L]}^{\oplus}\lambda {\mathbb{I}}\; d\lambda.
\end{equation}
\index{}
If the bias is present on the first lead, the situation is changed. Since $\mathcal{H}^{ac}(H_0)=
\mathcal{H}^{ac}(H^L)$, the generalized eigenfunctions of $H_0+vP_1$ are chosen to be
\begin{align}\label{prima2}
\Psi_1^{(v)}(\lambda;m)&:=\Psi(\lambda-v;m),\quad \lambda\in [-2t_L+v,2t_L+v],\quad m\geq 0,\\
\label{adoua2}\Psi_2^{(v)}(\lambda;m)&:=\Psi(\lambda;m),\quad \lambda\in [-2t_L,2t_L],\quad m\geq 0.
\end{align}
The corresponding generalized Fourier transformations are:
\begin{align}\label{genfour22}
&F_v\colon \bigoplus_{\alpha=1}^2 l^2(\mathbb{N}_{\alpha}) \to L^2([-2t_L+v,2t_L+v])
\oplus L^2([-2t_L,2t_L])\\
 [F_v(\Phi)]_1(\lambda)&=
\sum_{m\geq 0}\overline{\Psi(\lambda-v;m)}\Phi_1(m),\quad [F_v(\Phi)]_2(\lambda)=
\sum_{m\geq 0}\overline{\Psi(\lambda;m)}\Phi_2(m) \ .
\end{align}
Therefore, we can construct generalized eigenfunctions of $H+vP_1$, as solutions of the
Lippmann-Schwinger equation:
\begin{align}
\Phi_\alpha^{(v)}(\lambda;\cdot)= \Psi_\alpha^{(v)}(\lambda;\cdot)-
(H_0+vP_1-\lambda-i0_+)^{-1}H^T\Phi_\alpha^{(v)}
(\lambda;\cdot) \ .\label{asasea29}
\end{align}
These generalized eigenfunctions have the following very useful \textit{intertwining properties}
between the
subspaces of absolute continuity of the operators $H_0+vP_1$ and $H+vP_1$, which can be formally
written as:
\begin{align}\label{acincea23}
\Phi_\alpha^{(v)}(\lambda;\cdot)&=\Omega_{+}^{(v)}\Psi_\alpha^{(v)}(\lambda;\cdot) \ ,\\
\Psi_\alpha^{(v)}(\lambda;\cdot)&=\{\Omega_{+}^{(v)}\}^*\Phi_\alpha^{(v)}(\lambda;\cdot) \ .
\label{asasea27}
\end{align}

The scattering operator $S^{(v)}: \mathcal{H}^{ac}(H_0+vP_1)\mapsto
\mathcal{H}^{ac}(H_0+vP_1)$ is a unitary
map acting on $\mathcal{H}^{\rm ac}(H_0+vP_1)=\mathcal{H}^{\rm
  ac}(H_0)=\bigoplus_{\alpha=1}^2 l^2(\mathbb{N}_\alpha)$, and it is
given by $S^{(v)}=\{\Omega_{-}^{(v)}\}^*\Omega_+^{(v)}$. Then the
corresponding transition $T$-operator is
defined by $2\pi iT^{(v)}:= {\mathbb{I}}-S^{(v)}$. In the spectral
representation of $H^L+vP_1$ in the space $\int_{[-2t_L,2t_L]}^\oplus
\mathbb{C}^2d\lambda$, the $T$-operator is a $\lambda$-dependent $2\times
2$ matrix with elements denoted by
$t_{\alpha\beta}^{(v)}(\lambda)$. Using (\ref{genfouradj-b}) one gets the representation:
\begin{equation}\label{antcorn}
\sum_{\beta=1,2}t_{\alpha\beta}^{(v)}(\lambda)\Xi_\beta(\lambda)=
\frac{1}{2\pi i}[F({\mathbb{I}}-S^{(v)})F^*\Xi]_\alpha
(\lambda) \ .
\end{equation}
Then with the help of the generalized eigenfunctions we can express the $T$-matrix elements as:
\begin{equation}\label{antcorn22}
 t_{\alpha\beta}^{(v)}(\lambda):=\langle \Psi_\alpha^{(v)}(\lambda;\cdot),
 H^T \Phi_\beta^{(v)}(\lambda;\cdot)\rangle.
\end{equation}
Since $S$ is unitary, one gets the relation $i(T-T^*)=2\pi T^*T=2\pi TT^*$ (Optical Theorem),
which implies:
\begin{align}\label{anoua2}
 \imag\{t_{22}^{(v)}(\lambda)\}&=\pi\left (|t_{22}^{(v)}(\lambda)|^2+|t_{12}^{(v)}(\lambda)|^2\right ), \\
|t_{21}^{(v)}(\lambda)|^2&=|t_{12}^{(v)}(\lambda)|^2.\label{azecea2}
\end{align}

The \textit{transmittance} ${\cal T}_{\alpha\beta}^{(v)}(\lambda)$ between the leads $\alpha$
and $\beta$ for
a given energy $\lambda$ is defined by:
\begin{equation}\label{transmys}
{\cal T}_{\alpha\beta}^{(v)}(\lambda):= |t_{\alpha\beta}^{(v)}(\lambda)|^2 \ .
\end{equation}
Note that by definitions (\ref{prima2}), (\ref{adoua2}) and (\ref{antcorn22}) the transmittance
${\cal T}_{12}^{(v)}(\lambda)=0$ if $\lambda\not\in [-2t_L+v,2t_L+v]\cap [-2t_L,2t_L]$.

\subsection{Proof of {\rm (i)}}
By (\ref{prima1}) and (\ref{adoua1}) the evolution operator $U(t)$ obeys for $t>t_1$ the equation:
\begin{equation}\label{prima8}
 U(t)=e^{-i(t-t_1)(H+vP_1)}U(t_1) \ .
\end{equation}
Then by $\mathcal{H}= \mathcal{H}^{pp}(H+vP_1)\oplus \mathcal{H}^{ac}(H+vP_1)$ and by (\ref{atreia1})
we obtain for the current (\ref{acincea1}) measured at site $n$ the representation:
\begin{align}\label{adoua8}
I(t,n)&={\rm Tr} \{e^{-i(t-t_1)(H+vP_1)}U(t_1)f(H)U^*(t_1)e^{i(t-t_1)(H+vP_1)}j_n\}\nonumber \\
&={\rm Tr} \{e^{-i(t-t_1)(H+vP_1)}U(t_1)f(H)U^*(t_1)e^{i(t-t_1)(H+vP_1)}E_{\rm pp}(H+vP_1)j_n\}\nonumber \\
&+{\rm Tr} \{e^{-i(t-t_1)(H+vP_1)}U(t_1)f(H)U^*(t_1)e^{i(t-t_1)(H+vP_1)}E_{\rm ac}(H+vP_1)j_n\}\nonumber\\
&=:I_{\rm pp}(t,n)+I_{\rm ac}(t,n) \ ,
\end{align}
where $E_{\rm pp}(H+vP_1)$ denotes the projection on the pure point
subspace $\mathcal{H}^{pp}(H+vP_1)$.
By virtue of (\ref{apatra1}) one gets the identity:
\begin{align}\label{prima11}
&e^{i(t-t_1)(H+vP_1)}E_{\rm pp}(H+vP_1)j_ne^{-i(t-t_1)(H+vP_1)}\nonumber \\
&=\frac{d}{dt}\left
\{E_{\rm pp}(H+vP_1)e^{i(t-t_1)(H+vP_1)}P_2^{(n)}e^{-i(t-t_1)(H+vP_1)}\right\}.
\end{align}
Let us for now assume that $H+vP_1$ has a
finite number of eigenvalues. This means that $E_{\rm pp}(H+vP_1)$ is
trace class. Now if $T>t_1$, the pure point part of \eqref{adoua8} yields:
\begin{align}\label{atreia8}
\int_0^T I_{\rm pp}(t,n)dt &=\int_0^{t_1} I_{\rm pp}(t,n)dt + \int_{t_1}^{T} I_{\rm pp}(t,n)dt\nonumber
= \int_0^{t_1} I_{\rm pp}(t,n)dt \\
&+{\rm Tr} \left
\{U(t_1)f(H)U^*(t_1) \, D(T) \, E_{\rm pp}(H+vP_1)\right\} \ ,
\end{align}
where the operator
$D(T):=e^{i(T-t_1)(H+vP_1)}P_2^{(n)}e^{-i(T-t_1)(H+vP_1)} - P_2^{(n)}$ is
uniformly bounded in $T$. Since the first integral in the right-hand side of (\ref{atreia8})
is finite, the pure point spectrum does not contribute to the ergodic limit
\eqref{asasea1}.
In the case when $E_{\rm pp}(H+vP_1)$ does not have finite rank,
we have to employ an $\epsilon/2$ argument based on the fact that
$E_{\rm pp}(H+vP_1)j_n$ can be arbitrarily well approximated in the trace norm with an
operator containing the projection on a sufficiently large (but
finite) number of eigenvalues of $H+vP_1$. This approximation will be
independent of $T$, so the previous argument can be repeated.

So, it remains to investigate $I_{\rm ac}(t,n)$ and to show that it
actually converges when $t\to\infty$.
To this end we start with three technical lemmas:
\begin{lemma}\label{lema1}
The operators $U(t_1)-e^{-iH_0t_1-ivP_1\int_0^{t_1}\phi(\tau)d\tau}$ and $U^*(t_1)-e^{iH_0t_1+
ivP_1\int_0^{t_1}\phi(\tau)d\tau}$ are compact.
\end{lemma}
\begin{proof}
Since the following Dyson-type equation:
$$\frac{d}{dt}\left \{e^{iH_0t+ivP_1\int_0^{t}\phi(t)dt}U(t)\right \}=
-ie^{iH_0t+ivP_1\int_0^{t}\phi(\tau)d\tau}H^TU(t),$$
is equivalent to
$$U(t_1)=e^{-iH_0t_1-ivP_1\int_0^{t_1}\phi(\tau)d\tau}-i\int_0^{t_1}e^{-iH_0 (t_1-t)-ivP_1\int_t^{t_1}
\phi(\tau)d\tau} H^TU(t)dt \ ,$$
we use that $H^T$ is a compact (finite-rank) operator in order to finish the proof.
\end{proof}
\begin{lemma}\label{lema2}
 The operator $U(t_1)f(H)U^*(t_1)-f(H_0)$ is compact.
\end{lemma}
\begin{proof}
It is an easy consequence of Lemma \ref{lema1}, of the fact that $H_0$ commutes with $P_1$, and of the
observation that the difference $f(H)-f(H_0)$ is a compact (even trace-class) operator.
\end{proof}

\begin{lemma}\label{lema3}
 Let $K$ be a compact operator. Then the following trace-norm tends to zero:
\begin{equation}\label{adoua11}
 \lim_{t\to\infty}||Ke^{i(t-t_1)(H+vP_1)}E_{\rm ac}(H+vP_1)j_n||_1=0.
\end{equation}
\end{lemma}
\begin{proof}
Since $j_n$ is from the trace-class (finite rank in our case), by standard
$\epsilon/2$ arguments we can assume that operator $K$ has a finite rank. Then
the proof is a consequence of the Riemann-Lebesgue lemma.
\end{proof}

\begin{corollary}\label{corr-ac-current}
Use the identity $U(t_1)f(H)U^*(t_1)= (U(t_1)f(H)U^*(t_1)-f(H_0)) + f(H_0)$ in the representation of
$I_{\rm ac}(t,n)$. Then Lemma \ref{lema2} and Lemma \ref{lema3} imply the limit:
\begin{equation}\label{atreia11}
\lim_{t\to\infty}|I_{\rm ac}(t,n)-{\rm Tr} \{e^{-i(t-t_1)(H+vP_1)}f(H_0)e^{i(t-t_1)(H+vP_1)}E_{\rm ac}
(H+vP_1)j_n\}|=0.
\end{equation}
\end{corollary}

Now, to prove the ergodic limit \eqref{asasea1} it is enough to check
that the trace appearing in (\ref{atreia11}) converges when $t\to\infty$. To
this end we use a standard trick of inserting the free evolution and
then to use the identity:
\begin{equation*}
f(H_0)= e^{i(t-t_1)(H_0+vP_1)}f(H_0)e^{-i(t-t_1)(H_0+vP_1)}
\end{equation*}
in (\ref{atreia11}). Using \eqref{omegapm} together with the fact that the
wave operators are complete thus unitary, we
obtain the existence of the following strong limit:
$$\{\Omega_{+}^{(v)}\}^*E_{\rm ac}
(H+vP_1)=s-\lim_{t\to \infty}e^{-i(t-t_1)(H_0+vP_1)} e^{i(t-t_1)(H+vP_1)} E_{\rm ac}
(H+vP_1),$$
where the limit operator projects onto
$\mathcal{H}^{\rm ac}(H_0)$. Finally, because $j_n$ is trace class we
can conclude that the limit:
\begin{equation}\label{apatra11}
\lim_{t\to\infty}I_{\rm ac}(t,n)={\rm Tr} \left \{\Omega_{+}^{(v)}f(H^L)\{\Omega_{+}^{(v)}\}^*E_{\rm ac}
(H+vP_1)j_n\right \} \ ,
\end{equation}
exists and is finite.
\begin{remark}\label{H0-HL}
We were able to replace $f(H_0)=f(H^S)\oplus f(H^L)$ by $f(H^L)$ because the inner
sample is projected out by the wave operator $\{\Omega_{+}^{(v)}\}^*$ on the right.
\end{remark}
Until now we proved that the ergodic limit \eqref{asasea1} is
independent of $\phi$ and $t_1$. The independence
of $n$ follows from the next lemma:
\begin{lemma}\label{lemma4}
For any $n\geq 1$ one can establish the following continuity equation:
\begin{align}\label{acincea11}
{\rm Tr} \left \{\Omega_{+}^{(v)}f(H^L)\{\Omega_{+}^{(v)}\}^*E_{\rm ac}(H+vP_1)j_n\right \}&={\rm Tr} \left
\{\Omega_{+}^{(v)}f(H^L)\{\Omega_{+}^{(v)}\}^*E_{\rm ac}(H+vP_1)j_0\right \}\nonumber \\
&=:I_\infty.
\end{align}
\end{lemma}
\begin{proof}
Denote by $\chi_n := P_2 - P_{2}^{(n)}$ the projection on the first $n$ sites of the second lead.
Then \eqref{apatra1} and Remark \ref{supp-P} yield $j_0-j_n=i[H+vP_1,\chi_n]$. Hence, the identity
\eqref{acincea11} is equivalent to
$${\rm Tr} \left \{\Omega_{+}^{(v)}f(H^L)\{\Omega_{+}^{(v)}\}^*E_{\rm
    ac}(H+vP_1)[H+vP_1,\chi_n]\right \}=0.$$
But the operator $\chi_n$ is trace-class, so we can undo the
commutator. The wave operators intertwine between $H+vP_1$ and
$H_0+vP_1$, and $H_0+vP_1$ commutes with $H^L$. It follows that $H+vP_1$ commutes with
$\Omega_{+}^{(v)}f(H^L)\{\Omega_{+}^{(v)}\}^*E_{\rm ac}(H+vP_1)$. Then
the trace cyclicity finishes the proof of the lemma.
\end{proof}

\subsection{Proof of {\rm (ii)}}

We start by proving that the current at $t=0$ (i.e. at equilibrium) is zero for all $n$.
Indeed, according to \eqref{apatra1} and \eqref{acincea1} one has
$I(0,n)={\rm Tr}(f(H)j_n)=i{\rm Tr}(f(H)[H,P_2^{(n)}])$ for all $n\geq
0$. Now fix $n$ and denote by $H_N$ the
Dirichlet \textit{restriction} of the operator $H$ to the finite leads
of length $N <\infty$, where $n<N$.
Denote by $P_2^{(n),N}$ the finite-rank projection corresponding to
the restriction of $P_2^{(n)}$ to the bounded second lead. Then one
can prove a certain "thermodynamic limit" result \cite{CJM1}:
\begin{equation}\label{marrtie1}
I(0,n)=\lim_{N\to\infty} {\rm Tr}\left
  \{f(H_N)i[H_N,P_2^{(n),N}]\right \}.
\end{equation}
To understand why (\ref{marrtie1}) holds true, note that
$j_n=i[H_N,P_2^{(n),N}]=i[H,P_2^{(n)}]$ is a finite-rank operator
which is independent of $N$. Moreover, $f(H)$ and $f(H_N)$ differ
significantly from each other only very far from the support of
$j_n$. Details can be found in \cite{CJM1}.

But ${\rm Tr}\left \{f(H_N)i[H_N,P_2^{(n),N}]\right \}=0$
for all $N$ by trace cyclicity. Thus \eqref{marrtie1} shows that
$I(0,n)=0$ for any  $n\geq 0$.

The next step of the proof is to show that for all $t\geq 0$ one has:
\begin{equation}\label{marrtie2}
\lim_{n\to \infty}|I(t,n)-I(0,n)|=0 \ .
\end{equation}
Then by $I(0,n)=0$ for all $n$, the limit in \eqref{marrtie2} would
imply \eqref{martie101}. First we present
the difference in \eqref{marrtie2} as:
\begin{align}\label{marrtie3}
I(t,n)-I(0,n)={\rm Tr}\{U(t)f(H)[U^*(t)-e^{itH}]j_n\}+
{\rm Tr}\{[U(t)-e^{-itH}]f(H)e^{itH}j_n\}.
\end{align}
Then we express the propagator $U(t)$ (\ref{adoua1}) with the help of its time-ordered Dyson series:
\begin{align}\label{marrtie4}
&U(t)=e^{-itH}\\
&+e^{-itH}\sum_{k\geq 1}\frac{(-i)^kv^k}{k!}
\int_0^td\tau_1\int_0^{t}d\tau_2...\int_0^{t}d\tau_{k}\mathbb{T}\{
\phi(\tau_1)e^{i\tau_1H}P_1e^{-i\tau_1H}...\phi(\tau_{k})e^{i\tau_{k}H}P_1e^{-i\tau_{k}H}\}\nonumber
\end{align}
whereas its adjoint is given by:
\begin{align}\label{marrtie5}
&U^*(t)=e^{itH}\\
&+\sum_{k\geq 1}\frac{i^kv^k}{k!}
\int_0^td\tau_1\int_0^{t}d\tau_2...\int_0^{t}d\tau_{k}\mathbb{\widetilde{T}}\{
\phi(\tau_1)e^{i\tau_1H}P_1e^{-i\tau_1H}...\phi(\tau_{k})e^{i\tau_{k}H}P_1e^{-i\tau_{k}H}\}e^{itH} \ ,
\nonumber
\end{align}
where $\mathbb{T}$ means time-ordering in \textit{decreasing} order and
$\mathbb{\widetilde{T}}$ means time-ordering in \textit{increasing} order.

Note that in the formula \eqref{marrtie3} the first term on the
right-hand side contains the operator
$[U^*(t)-e^{itH}]j_n$; we want to show that its trace-norm goes to
zero with $n$. By a simple support
property (\ref{comm1-P}) one has $P_2^{(n-1)}j_n=j_n$ and since $j_n$
is a finite-rank, it is enough to prove
that $[U^*(t)-e^{itH}]P_2^{(n-1)}$ converges to zero with $n$ in the operator norm. To this end we need
a technical estimate given by the following lemma:
\begin{lemma}\label{lemadupa}
For any fixed $t\geq 0$ one has:
\begin{equation}\label{marrtie300}
\lim_{n\to \infty}\sup_{|\tau|\leq t}\left \Vert P_1 e^{i\tau H}P_2^{(n-1)}\right \Vert=0.
\end{equation}
\end{lemma}
\begin{proof}
Since the operator $H$ is bounded, for any $\epsilon>0$ there exists $N_\epsilon$ such that
\begin{equation}\label{marrtie6}
\sup_{|\tau|\leq t}\left \Vert e^{i\tau H}-\sum_{k=0}^{N_\epsilon}\frac{i^k\tau^k H^k}{k!}\right
\Vert\leq \epsilon.
\end{equation}
The support properties (Remark \ref{supp-P}) and the one-step hopping in the Hamiltonian $H^L$ imply
that $P_1 H^k P_2^{(n-1)}=0$ if $n > N_\epsilon \geq k$. Hence, by \eqref{marrtie6}
we obtain that for $n > N_\epsilon$
\begin{equation}\label{marrtie7}
\sup_{|\tau|\leq t}\left \Vert P_1e^{i\tau H}P_2^{(n-1)} \right \Vert\leq \epsilon \ ,
\end{equation}
which proves the lemma. \end{proof}

Applying this result to the expansion \eqref{marrtie5}, one finds that
$[U^*(t)-e^{itH}]P_2^{(n-1)}$ converges
to zero in norm. This convergence allows to bound from above
the limit of the difference \eqref{marrtie3}:
\begin{align}\label{marrtie8}
\limsup_{n\to \infty} |I(t,n)-I(0,n)|\leq \limsup_{n\to \infty}
|{\rm Tr}\{[U(t)-e^{-itH}]f(H)e^{itH}j_n\}|.
\end{align}

To estimate the limit \eqref{marrtie8} we use the representation $[U(t)-e^{-itH}]f(H)e^{itH}j_n =
[U(t)-e^{-itH}]\{(\mathbb{I}- P_2^{([n/2])}) + P_2^{([n/2])}\}f(H)e^{itH}j_n$. Since again the
function $f(H)e^{itH}$ can be approximated in operator norm by polynomials in $H$, we can apply to
$(\mathbb{I}- P_2^{([n/2])})f(H)e^{itH}j_n$ the same line of reasoning as in Lemma \ref{lemadupa}
to establish:
\begin{equation}\label{one}
\lim_{n\to \infty}\|(\mathbb{I}- P_2^{([n/2])})f(H)e^{itH}P_2^{(n-1)}\| = 0,
\end{equation}
since the distance between the supports of $\mathbb{I}- P_2^{([n/2])}$ and of $P_2^{(n-1)}$ tends
to infinity.
For the term $[U(t)-e^{-itH}]P_2^{([n/2])}f(H)e^{itH}j_n$ we use the representation
\eqref{marrtie4} and Lemma \ref{lemadupa}, which imply that the norm of $[U(t)-e^{-itH}]P_2^{([n/2])}$
goes to zero with $n$. Together with \eqref{one} this proves that the
limit of the right hand side of \eqref{marrtie8} equals zero, thus
\eqref{marrtie2} follows.

\subsection{Proof of {\rm (iii)}}

By virtue of \eqref{apatra11} and \eqref{acincea11} one has $\lim_{t\to\infty}I_{\rm ac}(t,n)= I_\infty$.
Therefore, it only remains to estimate the current $I_{\rm pp}(t,n)$. This gives by \eqref{adoua8}:
\begin{equation*}
\sup_{t\geq 0}|I_{\rm pp}(t,n)|\leq ||E_{\rm pp}(H+vP_1)j_n||_1 \ .
\end{equation*}
Note the right-hand side of this estimate can be made arbitrarily
small by increasing $n$, since we assumed that we have finitely many
eigenfunctions which are
necessarily localized near the sample $S$, thus
$$\lim_{n\to\infty}||E_{\rm pp}(H+vP_1)P_2^{(n-1)}||=0.$$
 This finishes the proof of (iii). Note that in the exceptional case
 in which $H+vP_1$ could have infinitely many eigenvalues, this
 argument fails.

\subsection{Proof of {\rm (iv)}}

To calculate the steady charge current \eqref{prima5} we use our main formula \eqref{acincea11} in the form:
\begin{equation*}
I_\infty = {\rm Tr} \left\{\Omega_{+}^{(v)}f(H^L)\{\Omega_{+}^{(v)}\}^*E_{\rm ac}(H+vP_1)j_0\right\} =
{\rm Tr} \left\{f(H^L)\{\Omega_{+}^{(v)}\}^*E_{\rm ac}(H+vP_1)j_0
\Omega_{+}^{(v)}\right\} \ .
\end{equation*}
Now using the spectral representation for $H^L+vP_1$ one can evaluate the
\textit{trace} on $l^2(\mathbb{N}_1)\oplus l^2(\mathbb{N}_2)$
with the help of its generalized eigenfunctions
(see \eqref{prima2} and \eqref{adoua2}). Then we obtain the representation:
\begin{align}
I_\infty &= \int_{-2t_L+v}^{2t_L+v} d\lambda \ f(\lambda-v)
\left\langle \Psi_1^{(v)}(\lambda;\cdot), \{\Omega_{+}^{(v)}\}^*
E_{\rm ac}(H+vP_1)j_0 \Omega_{+}^{(v)}\Psi_1^{(v)}(\lambda;\cdot)\right \rangle \nonumber \\
&+ \int_{-2t_L}^{2t_L} d\lambda \ f(\lambda)
\left\langle \Psi_2^{(v)}(\lambda;\cdot),\{\Omega_{+}^{(v)}\}^*
E_{\rm ac}(H+vP_1)j_0 \Omega_{+}^{(v)}\Psi_2^{(v)}(\lambda;\cdot)\right\rangle \ .     \label{curr1}
\end{align}

\noindent By \eqref{acincea23} for the scalar product in the first integral we get:
\begin{align}\label{asaptea11}
&i \left\langle \Psi_1^{(v)}(\lambda;\cdot), \{\Omega_{+}^{(v)}\}^*
[H^T,P_2] \Omega_{+}^{(v)}\Psi_1^{(v)}(\lambda;\cdot)\right \rangle
=2\imag{\left \langle \Phi_1^{(v)}(\lambda;\cdot),P_2 H^T \Phi_1^{(v)}(\lambda;\cdot)\right\rangle}
\nonumber \\
&=2\imag {\left \langle P_2\{\Psi_1^{(v)}(\lambda;\cdot)-(H_0+vP_1-\lambda-i0_+)^{-1}H^T\Phi_1^{(v)}
(\lambda;\cdot)\}, H^T\Phi_1^{(v)}(\lambda;\cdot)\right \rangle} \ ,
\end{align}
where in the second line we used the Lippmann-Schwinger equation \eqref{asasea29}.
Notice that the vector
$P_2 (H_0+vP_1-\lambda-i0_+)^{-1}H^T\Phi_1^{(v)}(\lambda;\cdot) \in
l^{2}(\mathbb{N}_{ 2})$ for almost every $\lambda$.
By Remark \ref{supp-P} one has: $[H_0 , P_2] = 0$ and $P_2 P_1 = 0$, which implies
$P_2 (H_0+vP_1-\lambda-i0_+)^{-1}H^T = P_2 (H_0 -\lambda-i0_+)^{-1}H^T$. Taking this and the identity:
$P_2\Psi_1^{(v)}(\lambda;\cdot)=0$ into account,  we can use the spectral
representation of $H_0$ and decomposition the vector
$P_2 (H_0 -\lambda-i0_+)^{-1}H^T \Phi_1^{(v)}(\lambda;\cdot)$
over the generalised eigenvectors
$\{\Psi_{2}^{(v)}(\lambda^{\prime};\cdot)\}_{\lambda^{\prime}\in[-2t_L,2t_L]}$
to obtain
\begin{align}
&2\imag {\left \langle P_2\{\Psi_1^{(v)}(\lambda;\cdot)-(H_0+vP_1-\lambda-i0_+)^{-1}H^T\Phi_1^{(v)}
(\lambda;\cdot)\}, H^T\Phi_1^{(v)}(\lambda;\cdot)\right \rangle} = \nonumber \\
&= -2\imag \int_{-2t_L}^{2t_L} d\lambda^{\prime} \frac{1}{\lambda^{\prime} -\lambda-i0_+}\left |\left
\langle\Psi_2^{(v)}(\lambda^{\prime};\cdot)H^T\Phi_1^{(v)}(\lambda;\cdot)\right \rangle\right |^2
\nonumber \\
&=-2\pi {\cal T}_{21}^{(v)}(\lambda) \ \chi_{[-2t_L,2t_L]}(\lambda) \ . \label{aopta11}
\end{align}
For the last equality we used the Sokhotskii-Plemelj formula, and definitions \eqref{antcorn22},
\eqref{transmys}.

By the same line of reasoning one gets for the second integrand in \eqref{curr1}:
\begin{align}\label{anoua11}
&\left\langle \Psi_2^{(v)}(\lambda;\cdot),\{\Omega_{+}^{(v)}\}^*
E_{\rm ac}(H+vP_1)j_0 \Omega_{+}^{(v)}\Psi_2^{(v)}(\lambda;\cdot)\right\rangle
=2\imag{\left \langle \Phi_2^{(v)}(\lambda;\cdot),P_2 H^T\Phi_2^{(v)}(\lambda;\cdot)\right \rangle}
\nonumber \\
&=2\imag {\left \langle
P_2\{\Psi_2^{(v)}(\lambda;\cdot)-(H_0+vP_1-\lambda-i0_+)^{-1}H^T\Phi_2^{(v)}(\lambda;\cdot)\}, H^T
\Phi_2^{(v)}(\lambda;\cdot)\right \rangle}\nonumber \\
&=2\imag \{t_{22}^{(v)}(\lambda)\}-
2\imag {\left \langle P_2(H_0-\lambda-i0_+)^{-1}H^T\Phi_2^{(v)}(\lambda;\cdot),
H^T\Phi_2^{(v)}(\lambda;\cdot)\right \rangle}\nonumber \\
&=2\imag \{t_{22}^{(v)}(\lambda)\}-2\pi |t_{22}^{(v)}(\lambda)|^2 = 2\pi {\cal T}_{12}^{(v)}(\lambda) \ ,
\end{align}
where for the last identity we used \eqref{antcorn22} and \eqref{anoua2}. Note that here
$\lambda\in [-2t_L,2t_L]$.
Taking into account the symmetry \eqref{azecea2} and plugging \eqref{aopta11}, \eqref{anoua11}
into \eqref{curr1}, we obtain \eqref{prima5}.

Recall that ${\cal T}_{12}^{(v)}(\lambda)=0$, if $\lambda\not \in [-2t_L,2t_L]\cap
[-2t_L+v,2t_L+v]$.   \qed

\section{{Concluding remarks}}

In the present paper, we established a Landauer-type formula for the
stationary current running through a
discrete system with a (small) sample coupled to one-dimensional
infinite leads. We give a rigorous proof of
the existence of the ergodic limit of the charge current and then its
explicit expression. Our strategy is based
on the partition-free approach, it is quite general and demands a
minimal information about the sample.

There are several open problems which deserve to be mentioned.

\begin{enumerate}

\item  One of
them is our Conjecture \ref{Lieb-Robinson}
about the Lieb-Robinson type correlation group velocity bound, which
up to our knowledge it has not been studied before in this context.

\item If $V(t)$ is a time dependent bias between which after $t=t_1$
  becomes a perfect monochromatic signal like $V_0+V_1\cos(\omega t)$, then
the ergodic limit exists and is independent of $t_1$
and of the site where one measures the current. Furthermore, the
ergodic limit is given by a Landauer-like formula \cite{Koh}.

\item A computation of the current $I(t,0)$ (see \eqref{acincea1}), by
  expressing the evolution unitaries through the functional calculus
  associated to the resolvents, and the
  resolvents with the help of the Feshbach formula as in
  \cite{CJM1, CJM2}. Can one obtain an "easy" formula for
  $I(t,n)$ at a given $n$? Can one study numerically the transient effects
  and check point (ii) in Theorem \ref{teorema1}?

\item Study the resonant transport in the case of small coupling
  ($0<\tau<<1$ and $v$ a variable parameter).

\item What happens with point {\it (ii)} of our theorem if there are infinitely
  many eigenvalues?

\item Study the "wide band limit", or $t_L\to\infty$.

\item Compute the first few corrections in $v$ of the conductivity
  tensor.

\item Introduce a Kohn-Sham interaction in the sample, as in
  Stefanucci's papers \cite{KSA}. How can one properly formulate the mathematical problem in
  this non-linear case? Can one still prove the existence of a steady
  state? Is it unique?

\end{enumerate}

\section{Acknowledgments}

This paper is dedicated to our late friend Pierre Duclos. In his large 
spectrum of interests the mathematical study of transient and steady-state currents 
took a special place including this project. We missed him working on this paper. 

\noindent We are grateful to G. Stefanucci for his valuable comments.
H.C. acknowledges support from the Danish FNU grant
{\it Mathematical Physics}.

\noindent Part of this work was carried out when H.C. and V.Z. were visiting the Bernoulli Center,
EPFL, Lausanne in February 2010.


\end{document}